\documentclass[aps,prd,twocolumn,nofootinbib]{revtex4-1}

\usepackage{graphicx}
\usepackage{color}

\newcommand{\uestc}{\affiliation{School of Physics, University of Electronic Science and
Technology of China, Chengdu 610054, China}}

\newcommand{\huzhou}{\affiliation{Yangtze Delta Region Institute (Huzhou), University of Electronic Science and Technology of China, Huzhou 313001, China}}

\newcommand{\cqu}{\affiliation{Department of Physics, Chongqing Key Laboratory for Strongly Coupled Physics, Chongqing University, Chongqing 401331, China}}

\begin{document}

\title{BEEC2.0: An Upgraded Version for the Production of Heavy Quarkonium at Electron-positron Collider}

\author{Zhi Yang}\email{zhiyang@uestc.edu.cn}
\huzhou
\uestc
\author{Xu-Chang Zheng}\email{zhengxc@cqu.edu.cn}
\cqu
\author{Xing-Gang Wu}\email{wuxg@cqu.edu.cn}
\cqu

\begin{abstract}

The event generator BEEC [Z. Yang, X.G. Wu and X.Y. Wang, Comput. Phys. Commun. {\bf 184}, 2848 (2013)] was devoted to the simulation of heavy quarkonium production at an unpolarized electron-positron collider. We upgraded it here by adding the generation of quarkonium with polarized electron and positron beams. In addition, the production of color-singlet 2$S$-wave states were included. Several future electron-positron colliders with high luminosity have been under discussion in the past decade. Especially, their possibility of producing polarized beams is important for providing more insights into the underlying physics. This upgraded version offers a useful tool for the feasibility study on quarkonium from the experimental side. \\

\noindent {\it Keywords:} Event generator, heavy quarkonium, electron-positron collider \\

\end{abstract}

\maketitle

\noindent{\bf NEW VERSION PROGRAM SUMMARY}\\

\noindent{\it Program title} : BEEC version 2.0\\

\noindent{\it CPC Library link to program files} : (to be added by Technical Editor) \\

\noindent{\it Code Ocean capsule} : (to be added by Technical Editor) \\

\noindent{\it Licensing provisions} : GNU General Public License 3 (GPL) \\

\noindent{\it Programming language} : FORTRAN 77/90 \\

\noindent{\it Journal Reference of previous version} : Comput. Phys. Commun. {\bf 184}, 2848 (2013)\\

\noindent{\it Does the new version supersede the previous version?} : Yes\\

\noindent{\it Reasons for the new version} : The proposed electron-positron colliders could possibly produce polarized beams, with which it would be helpful to provide more insights into the structure of the underlying physics in heavy quarkonium. Thus, it is essential to simulate the quarkonium production through polarized beams. In addition,  the $B_c(2S)$ state attracted more and more interests since its discovery, which makes it necessary to produce the 2$S$-wave excited states. \\

\noindent{\it Summary of revisions} : Two polarized amplitudes of leptonic scattering are computed theoretically. The relevant codes of the formula are included in BEEC2.0. Meanwhile, new parameters for the 2$S$-excited states are also included. \\

\noindent{\it Nature of problem} :  Although PYTHIA can generate events for many processes, a dedicated event generator is also necessary. Especially, PYTHIA does not internally produce events with polarized beams. BEEC is implemented into PYTHIA as an external process for simulating heavy quarkonium events with high efficiency. Further hadronization and decay simulation can be done by using PYTHIA subroutine. \\

\noindent{\it Solution method} : Two projection operators are used to calculate the polarized leptonic scattering amplitudes with the help of improved trace technology~\cite{Chang:1992bb,Chang:2007si}. The code with option can generate events through unpolarized and polarized initial beams. In addition to the ($Q\bar{Q'}$)-quarkonium $(Q, Q'=b, c)$ in color-singlet 1$S$-wave states, 1$P$-wave states, and the color-octet 1$S$-wave states in the last version, BEEC 2.0 also deal with the production of 2$S$-wave excited states within the framework of non-relativistic QCD~\cite{Bodwin:1994jh}. \\

\noindent{\it Additional comments including Restrictions and Unusual features} : About the running time, it depends on which option one chooses to match PYTHIA when generating the heavy quarkonium events. Typically, for the production of the $S$-wave quarkonium states, if setting IDWTUP=$1$ (unweighted events), then it takes about 21 minutes on a 3.22 GHz Apple M1 Pro Processor machine to generate $10^{5}$ events; if setting IDWTUP=$3$ (weighted events), it takes only $\sim5$ minutes to generate $10^5$ events. While for the production of the $P$-wave quarkonium states, the time will be almost one hundred times longer than the case of the $S$-wave quarkonium. \\

\section{INTRODUCTION}
The doubly heavy quark-antiquark bound state has been attracted lots of interests because of its special features that involve both the perturbative QCD and the associated non-perturbative physics~\cite{Brambilla:2010cs}. It provides a good platform to deepen our understanding on these two aspects.  Especially, for the $B_c$ meson, it only decay through weak interactions since it carries explicit flavors, which would be helpful to investigate the potential models as well as the weak decay mechanism. At the hadron colliders, the heavy quarkonium has been widely studied from both theoretical and experimental sides.  In particular, the dedicated event generator BCVEGPY was designed to simulate the production of the $B_c$ meson through hadron collision~\cite{Chang:2003cq,Chang:2005hq,Chang:2006xka,Wang:2012ah}, which has been widely used in the experimental analysis. 

Due to the cleaner background compared to hadron collider, the measurement in electron-positron collider can be made with much better precision. This is crucial for the comparison with theoretically investigation because the information of this heavy-flavour bound state can be precisely predicted. Several electron-positron colliders were proposed with luminosity up to ${\cal L}\propto 10^{34-36}cm^{-2}s^{-1}$, i.e., the super $Z$ factor~\cite{wjw}, the International Linear Collider (ILC)~\cite{Erler:2000jg}. For the sake of studies on experimental feasibility, the event generator, named as BEEC, was programmed to simulate the production of heavy quarkonium at an unpolarized electron-positron collider~\cite{Yang:2013vba}. The production rate is calculated in the framework of the effective theory of non-relativistic QCD (NRQCD)~\cite{Bodwin:1994jh}. In NRQCD, the differential cross section of the quarkonium production process, $e^{+}+e^{-} \rightarrow \left|\left(Q Q^{\prime}\right)[n]\right\rangle+Q^{\prime}+\bar{Q}$ with $Q$ and $Q^{\prime}$ stands for $b$ or $c$, can be factorized into
\begin{equation} 
d \sigma=\sum_{n} d \hat{\sigma}\left(e^{+}+e^{-} \rightarrow\left(Q \bar{Q}^{\prime}\right)[n]+Q^{\prime}+\bar{Q}\right)\left\langle\mathcal{O}^{H}(n)\right\rangle, \label{eq:nrqcd}
\end{equation}
where $\left\langle\mathcal{O}^{H}(n)\right\rangle$ and $d \hat{\sigma}$ are the non-perturbative matrix element and the differential short-distance cross section, respectively. The universal non-perturbative matrix element is proportional to the inclusive transition probability of the perturbative state $\left(Q Q^{\prime}\right)[n]$ into the bound state $\left|\left(Q Q^{\prime}\right)[n]\right\rangle$. 
The matrix elements can be determined by global fitting of the experimental data or directly related to the wave functions at the zero point (or the derivative of the wave function at the zero point) derived from certain potential models for the color-singlet case, some potential models can be found in Refs.~\cite{Eichten:1994gt,Li:2020kyb}. To deal with the short-distance cross section, the so-called ‘improved trace technology’~\cite{Chang:1992bb,Chang:2007si} was employed. Compared to the conventional squared amplitude approach, the improved trace technology can do the numerical calculation at the amplitude level, and hence improve the simulation efficiency greatly.

Recently, there are more discussions on the proposal of the future electron–positron colliders. Especially, these colliders could possibly produce polarized beams, for example the ILC~\cite{Moortgat-Pick:2005jsx}, the Future Circular Collider (FCC-ee)~\cite{FCC:2018evy} and the Circular Electron–Positron Collider (CEPC)~\cite{cepc}. Using polarized beams, it would provide more useful insights into the structure of the underlying physics in heavy quarkonium. Therefore, the simulation of the quarkonium production through polarized beams is necessary for the study at electron-positron colliders. Furthermore, after the discovery of $B_c(2S)$ state by ATLAS collaboration~\cite{ATLAS:2014lga}, it attracted more and more interests. The simulation of such 2$S$-excited state is also important.

In this paper, we upgrade the BEEC by including the relevant codes for producing the event with polarized initial beams, and for producing the 2$S$-excited states of heavy quarkonium. The  paper is organized as follows. In Sec.~\ref{sec:2}, we briefly review the last version of BEEC and illustrate the updates in detail. The final section is reserved for a short summary.

\section{The upgraded generator}\label{sec:2}
The BEEC is written in a PYTHIA-compatible formate in order to conveniently apply all PYTHIA~\cite{Sjostrand:2006za} subroutines. After the heavy quarkonium is generated, one can use the PYTHIA subroutines to read the information of final states, and do further simulation of hadronization and decay. The codes of quarkonium generation are systematically organized in seven modules according to their purpose.
\begin{enumerate}
\item The module \textbf{generate}: Its purpose is to initialize input parameters for event simulation, to calculate the kernel for the phase-space and do the integration. This is the key
module that generates the final events with the help of \textbf{coefficient}, \textbf{amplitude}, and \textbf{phase}. After the events are obtained, it establishs the connection between BEEC and PYTHIA.
\item The module \textbf{phase}: This module generates the phase-space points by using subroutine RAMBOS~\cite{Kleiss:1985gy} and transform the obtained four-momentum of the final particles to the module \textbf{generate}. Besides, in a grade file, it records the importance sampling produced by using subroutine VEGAS~\cite{Lepage:1977sw}.
\item The module \textbf{pybook}: Its purpose is to initialize the PYTHIA to use its subroutine PYBOOK for recording useful information during the simulation.
\item The module \textbf{setparameter}: The input parameters are simplified/optimized in this module. If one input parameter is out of its allowed range, this module will generate some error message and stop the program.
\item The module \textbf{system}: This module contains certain running messages, which will be displayed at the intermediate steps.
\item The module \textbf{coefficient}: This module stores the coefficients of all independent Lorentz-structures for the corresponding quarkonium states.
\item The module \textbf{amplitude}: This module uses the coefficients of Lorentz-structures stored in module \textbf{coefficient} to calculate the amplitude numerically according to the simplified and compact analytical formulas.
\end{enumerate}
All the modules are under the main code directory together with an input file input.dat and a file parameter.F reading the inputs. More details for the modules can be found in the first version of BEEC~\cite{Yang:2013vba}.

In order to fulfill the new task mentioned above, some changes will be made in the generator. We explain the main changes in the following:

\begin{itemize}
\item {\bf Polarized initial electron/positron}.

	For the production of heavy quarkonium, the scattering amplitude can be written as
	\begin{equation} 
	i{\cal M}={\cal C}\times L_{rr^{\prime}}^{\mu}D_{\mu\nu}H_{ss^{\prime}}^{\nu}, \label{eq:amp}
	\end{equation}
	where ${\cal C}$ is an overall parameter and $D_{\mu\nu}$ is the propagator of virtual photon. $L_{rr^{\prime}}^{\mu}$ and $H_{ss^{\prime}}^{\nu}$ are the leptonic and hadronic parts, respectively. The indices $r$, $r^{\prime}$, $s$ and $s^{\prime}$ denote the spin projections of the initial electron and positron, and final quark and antiquark, respectively. Thus the short-distance differential cross section in Eq.~(\ref{eq:nrqcd}) is
	\begin{equation}
	d \hat\sigma=\frac{1}{4 \sqrt{\left(p_{1} \cdot p_{2}\right)^{2}-m_{1}^{2} m_{2}^{2}}} \bar{\sum}|\mathcal{M}|^{2} d \Phi_{3},
	\end{equation}
	where $p_i$ and $m_i$ ($i=1,2$) are the momentums and masses of initial electron and positron, respectively. $\bar{\sum}$ stands for the sum over the color and spin of all final particles. $d \Phi_{3}$ is the three-particle phase space
	\begin{equation}
	d \Phi_{3}=(2 \pi)^{4} \delta^{4}\left(p_{1}+p_{2}-\sum_{f}^{3} q_{f}\right) \prod_{f=1}^{3} \frac{d^{3} q_{f}}{(2 \pi)^{3} 2 q_{f}^{0}}.
	\end{equation}
As mentioned in the introduction, the amplitude in Eq.~(\ref{eq:amp}) is calculated numerically by using the improved trace technology, with which the coefficients of Lorentz-structures and the compact analytical formulas can be obtained and are stored in modules \textbf{coefficient} and \textbf{amplitude}, respectively. More details about this method can be found in Ref.~\cite{Yang:2011ps}. Using this method, BEEC first generates proper phase-space points, and then calculates the numerical value for the amplitudes.
The squared-amplitudes averaged over the spin states of initial particles are included in the file sqamp.F of module {\bf amplitude}.

In order to calculate the polarized amplitudes, we can use the helicity states of the initial electron and positron to describe their polarized state. Thus we use the projection operators
\begin{equation}
P_{L}=\frac{1-\gamma^{5}}{2}, \quad P_{R}=\frac{1+\gamma^{5}}{2}
\end{equation}
to the leptonic part of the scattering amplitude
\begin{equation}
L_{rr^{\prime}}^{\mu} = \bar{v}_{r}\left(p_{2}\right) \Gamma^{\mu}u_{r^{\prime}}\left(p_{1}\right)
\end{equation}
with the vertex $\Gamma^{\mu}=\gamma^{\mu}\left(1-4 \sin ^{2} \theta_{w}-\gamma^5\right).$
Due to the small electron and positron masses, only two polarization cases are dominant in the production, which are $e^+_R e^-_L$ and $e^+_L e^-_R$ and can be respectively obtained as
\begin{equation}
\bar{v}_{r}\left(p_{2}\right) \Gamma^{\mu} P_{L} u_{r^{\prime}}\left(p_{1}\right)
\end{equation}
and
\begin{equation}
 \bar{v}_{r}\left(p_{2}\right) \Gamma^{\mu} P_{R} u_{r^{\prime}}\left(p_{1}\right).
\end{equation}
These two leptonic scattering amplitudes can be easily calculated by using the improved trace technology. In this upgraded generator, we add the corresponding coefficients of Lorentz-structures in module \textbf{coefficient}. Then we give one choice to calculate the squared-amplitude for different spin state and add one parameter to select the needed spin states. This flag parameter is included in relevant code files and the input file input.dat:
	
\begin{itemize}
\item {ipolar: determining which spin state of the initial particle should be generated. When ipolar=0 is set in the input file, the code file sqamp.F will calculate the squared amplitude for the unpolarized initial states. While for $\text{ipolar}=1,\;2$, the production for $e^+_R e^-_L$ and $e^+_L e^-_R$ polarization cases will be generated, respectively.}
\end{itemize}

In addition, two new parameters are included in the file input.dat:

\begin{itemize}
\item {pem, pep: denote the polarization of electron and positron beams, respectively. The default values are set to be 0.8 and can be changed according to specific collider. These two parameters are only used in the generation of unweighted event.}
\end{itemize}

We show in Fig.~\ref{fig:diff-pol} the differential cross section $d\sigma/d\text{cos}\theta$ for the production $e^+_L e^-_R \to B_c(^1S_0)+b+\bar{c}$, where $\theta$ is the angle between the momenta of the meson in final state and the electron in initial state. This distribution is consistent with that in the theoretical investigation.~\cite{Zheng:2015ixa}.

\begin{figure}[t]
\centering
\includegraphics[width=0.8\linewidth]{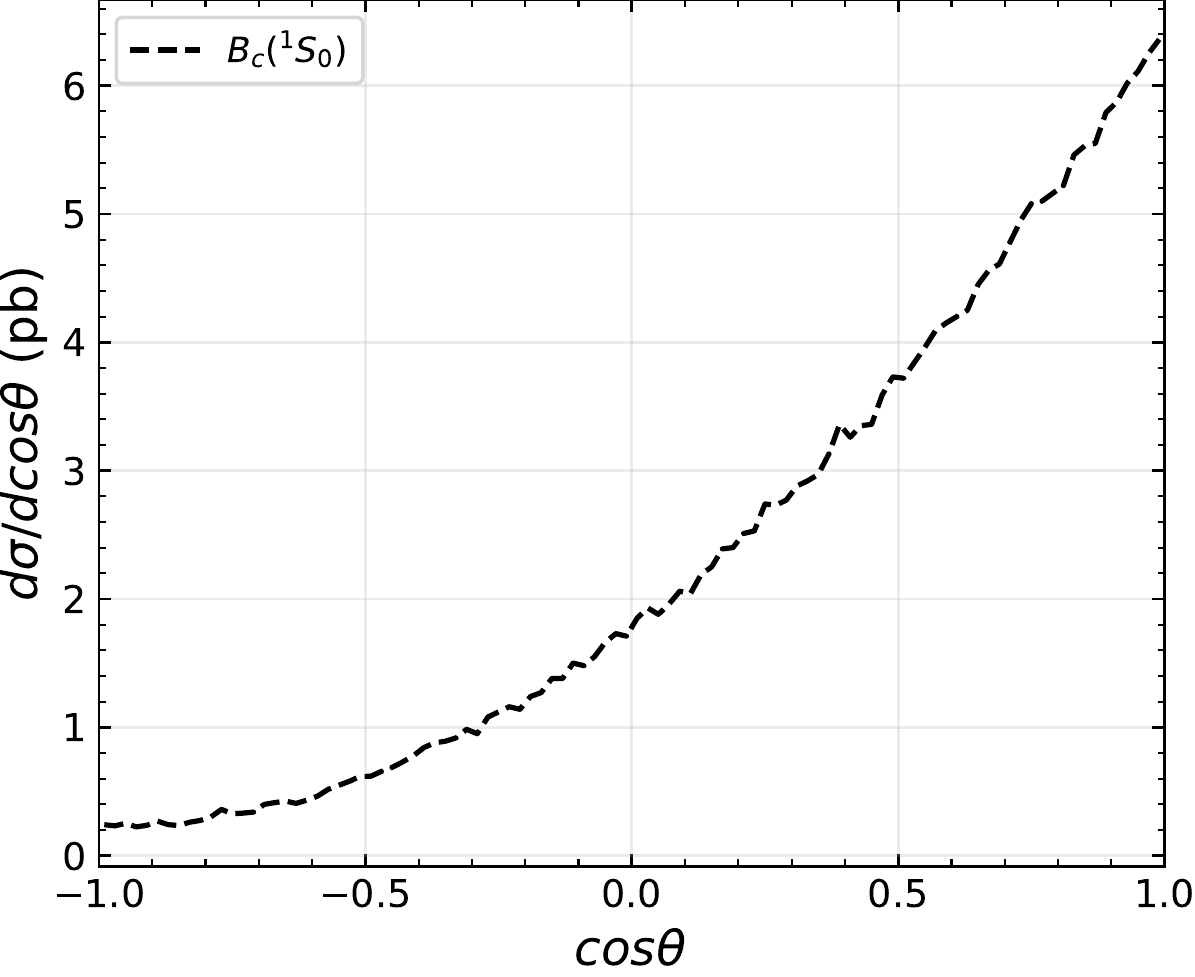}
\caption{The differential cross section $d\sigma/d\text{cos}\theta$ for the production $e^+_L e^-_R \to B_c(^1S_0)+b+\bar{c}$ with polarized initial beams.}
\label{fig:diff-pol}
\end{figure}

\item {\bf Generation of 2$S$-excited states}.
	
	Besides the production of $1S$-wave and $1P$-wave states, the $2S$-wave states are also involved in this version. The production amplitudes of $2S$-wave states have the same formula as that of $1S$-wave states, but with different $b$-quark mass {\bf pmb}, $c$-quark mass {\bf pmc}, and the radial wave function at the zero {\bf fbc}. Therefore, we can get the production of the $2S$-wave states by replacing the parameters of the $1S$-wave states with those of the $2S$-wave states. These parameters of the $2S$-wave states were shown in Tab.~\ref{tab:paras} for the three kinds of $Q\bar Q^{\prime}$ bound states.
	
\begin{table}[t]
\caption{The default values of the parameters for $2S$-wave states~\cite{Eichten:1994gt}. These values can be changed in the file parameter.F.}
\label{tab:paras}
\begin{ruledtabular}
\begin{tabular}{rccccc}

                                        &    $c\bar{c}$       &  $c\bar{b}$  & $b\bar{b}$\\
\hline
mass  (GeV)                    &   3.608                &   6.867       &  9.963 \\
{\bf fbc}  (GeV$^{3/2}$)   &   0.727                &   0.991       &  1.798 \\
{\bf pmb}  (GeV)              &      -                    &    5.234       &  4.981 \\
{\bf pmc}  (GeV)              &   1.804                &   1.633        & - \\

\end{tabular}
\end{ruledtabular}
\end{table}
	
	We added two new values for {\bf ibcstate} in this new version: {\bf ibcstate}=9 is for the generation of $2^1S_0$ state and {\bf ibcstate}=10 is for the generation of $2^3S_1$ state. While for the production of mixed events, the values for IMIXTYPE need to be rearranged. IMIXTYPE=1 is for the generation of the mixing events for all the color-singlet and color-octet $1S$-wave states, color-singlet $1P$-wave, and color-singlet  $2S$-wave states. IMIXTYPE=2 is for the generation of the mixing events for the $1^1S_0$ and $1^3S_1$ states. IMIXTYPE=3 is for the generation of the mixing events for the four $1P$-wave states plus the two color-octet $1^1S_0$ and $1^3S_1$ states. IMIXTYPE=4 is for the generation of the mixing events for the $2^1S_0$ and $2^3S_1$ states.
	\\
\end{itemize}

\section{Summary}

The event generator BEEC for simulating the production of heavy quarkonium through electron-positron annihilation is upgraded by including the production through polarized beams and color-singlet 2$S$-wave excited states. Considering the possible ability of producing polarized beams on the proposed electron-positron colliders, such upgrade would be useful for the feasibility studies on these colliders.

\begin{acknowledgments}
This work was supported in part by the Natural Science Foundation of China under Grant No.11847301 and No.12175025, and by the Fundamental Research Funds for the Central Universities under Grant No. 2019CDJDWL0005.
\end{acknowledgments}

\bibliography{beec2.bib}

\end{document}